\documentclass[12pt]{iopart}

\usepackage{cite}
\usepackage{graphicx}
\usepackage{iopams}
\usepackage{mathptmx}
\usepackage{textcomp}
\usepackage[english]{babel}
\usepackage{pslatex}
\usepackage{color}
\usepackage[normalem]{ulem}

\begin{document}

\title{
 Ab Initio Molecular Dynamics on the Electronic Boltzmann Equilibrium Distribution
}

\author{J. L. Alonso$^{1,2}$, A. Castro$^2$, P. Echenique$^{1,2,3}$,
  V. Polo$^{2,4}$, A. Rubio$^5$ and
  D. Zueco$^6$}
\address{$^1$ Departamento de F{\'{\i}}sica Te{\'{o}}rica, Universidad de Zaragoza, Pedro Cerbuna 12, E-50009 Zaragoza, Spain}
\address{$^2$Institute for Biocomputation and Physics of Complex Systems (BIFI), University of Zaragoza, Mariano
Esquillor s/n, E-50018 Zaragoza, Spain}
\address{$^3$Instituto de Qu{\'{\i}}mica F{\'{\i}}sica ``Rocasolano'', CSIC, Serrano 119, E-28006 Madrid, Spain}
\address{$^4$Departamento de Qu{\'{\i}}mica Org{\'{a}}nica y Qu{\'{\i}}mica F{\'{\i}}sica, Facultad de Ciencias, Universidad de Zaragoza, E-50009 (Spain)}
\address{$^5$Nano-Bio Spectroscopy group and ETSF Scientific Development Centre, Departamento de F{\'{\i}}sica de Materiales,
Universidad del Pa{\'{\i}}s Vasco, Centro de F{\'{\i}}sica de Materiales CSIC-UPV/EHU-MPC and DIPC
E-20018 San Sebasti{\'{a}}n, Spain}
\address{$^6$Instituto de Ciencia de Materiales de Arag{\'{o}}n and Departamento de F{\'{\i}}sica de la Materia Condensada
CSIC-Universidad de Zaragoza, E-50009 Zaragoza, Spain}
\ead{dzueco@unizar.es}


\begin{abstract}
  We prove that for a combined system of classical and quantum particles, it
  is possible to write a dynamics for the classical particles that
  incorporates in a natural way the Boltzmann equilibrium population
  for the quantum subsystem.  In addition, these molecular dynamics do
  not need to assume that the electrons immediately follow the nuclear
  motion (in contrast to any adiabatic approach), and do not present
  problems in the presence of crossing points between different
  potential energy surfaces (conical intersections or
  spin-crossings). A practical application of this molecular dynamics
  to study the effect of temperature in molecular systems presenting
  (nearly) degenerate states -- such as the avoided crossing in the
  ring-closure process of ozone -- is presented.
\end{abstract}

\maketitle

\section{Introduction}

The possibility of dividing a system of particles into a
\emph{quantum} and a \emph{classical} subsystem is of wide interest in
several areas of physics and chemistry. This division is typically,
but not exclusively, done by considering the electrons to be quantum,
and the nuclei to be classical (this is the choice that we make in
this work, although the observation of quantum effects for protons is
nowadays a matter of considerable debate~\cite{bothma-2009}).  When
the gap is large (i.e. when the lowest electronic excitation energy is
much greater than the highest available nuclear vibrational frequency,
which depends on the temperature), the division is properly handled by
making use of the standard Born-Oppenheimer (BO) separation. Ab-initio
Molecular Dynamics (AIMD) \cite{aimd} can then be performed for the
system of classical nuclei on the ground state BO surface (gsBOMD) --
and this type of dynamics has been applied countless times in the last
couple of decades, either directly or by making use of equivalent but
more efficient schemes such as the Car-Parrinello (CP)
technique~\cite{car-1985} or other alternatives~\cite{alonso-2008,
  kuhne-2007}. These dynamics can be used to compute equilibrium
averages by assuming ergodicity and computing time averages over a
number of trajectories. Notice that when the molecular dynamics is
used to sample the equilibrium distribution, all the dynamics are
physically equivalent insofar as they produce the correct
distribution, and only efficienty criteria may favour one over
another.

If the temperature is of the order of the electronic gap or larger we
cannot assume any longer that the quantum particles are constantly in
the ground state.  Indeed, in many physical, chemical or biological
processes the dynamical effects arising from the presence of low lying
electronic excited states have to be taken into account. For instance,
in situations where the Hydrogen bond is weak, different states come
close to each other and non-adiabatic proton transfer transitions
become rather likely at normal temperature~\cite{may-2004}. In these
circumstances, the computation of ensemble averages cannot be based on
a model that assumes the nuclei moving on the ground state BO surface.

If transitions to higher energy
levels must be allowed, a different type of dynamics must be
performed. In the realm of first principles MD calculations, two
approaches come to mind: Ehrenfest dynamics, and surface
hopping~\cite{tully-1998}.  Their suitability for the calculation of
equilibrium properties is however still a subject of
study~\cite{mauri-1993, parandekar-2005, schmidt-2008, bastida-2007}.
Also, in density-functional theory (DFT), one could perform MD at
$T\ne0$, working with partial occupation numbers to account for the
electronic excitations~\cite{tdependentdft}, ideally making use of a
temperature-dependent exchange and correlation
functional~\cite{mermin}. This scheme is however tied to DFT, and is
hindered by the difficulty of realistically approximating this
functional.

In this work, we first emphasize that the distribution that is often
regarded in the context of quantum-classical systems as the ``correct
equilibrium distribution'' ($\rho^{\rm eq \, (0)}_W $ in   Eq.~(\ref{eq:qcrho})
below), despite being commonly obtained through heuristic arguments,
is however only a zero-th order approximation (in the
quantum-classical mass ratio $\sqrt {m/M}$) to the canonical
equilibrium density matrix associated to a rigorous quantum-classical
formulation, as shown by Nielsen, Kapral and
Ciccotti~\cite{kapral-1999, nielsen-2001}.  Therefore, there is a
priori no reason to ask for any mixed quantum-classical theory such
as, e.g. Ehrenfest dynamics, to yield exactly the Boltzmann
equilibrium distribution, $\rho^{\rm eq \,(0)}_W $, contrarily to
what is often required in the literature~\cite{mauri-1993,
  parandekar-2005, schmidt-2008, bastida-2007}.

However, although $\rho^{\rm eq \, (0)}_W $ is only an approximation to the
correct quantum-classical equilibrium ensemble, it is acceptable when $\sqrt
{m/M}$ is small, and the results are then reliable. Hence, $\rho^{\rm eq \, (0)}_W $ will also be the target equilibrium distribution in this work.

Therefore, in what follows, we will write a system of dynamic
equations for the classical particles such that the equilibrium
distribution in the space of classical variables is in fact given by
Eq.~(\ref{eq:qcrho}). This is also a goal of surface hopping methods,
although it is not fully achieved since these methods do not exactly
yield this distribution.  We will do this by deriving a temperature
dependent effective potential for the classical variables, which
differs from the potential energy surfaces (PES) that emerge from BO
equations. It is straightforward, however, to write an equation that
gives the expression for the effective potential in terms of these
PES. But note that our approach will be based on the assumption that
the full system of electrons and nuclei is in thermal equilibrium at a
given temperature, and  not on the assumption that electrons immediately
follow the nuclear motion (i.e. the adiabatic approximation).

As an example of the interest in going beyond the PES we mention the
issue of quantum effects in proton transfer~\cite{iyengar-2008}. It is
a matter of current debate to what extent protons behave
"quantum-like" in biomolecular systems (e.g.  is there any trace of
superposition, tunneling or entanglement in their
behavior?). Recently, McKenzie and coworkers~\cite{bothma-2009} have
carefully examined the issue, and concluded that ``tunneling well
below the barrier only occurs for temperatures less than a temperature
$T_0$ which is determined by the curvature of the PES at the top of
the barrier.'' In consequence, the correct determination of this
curvature is of paramount importance. 

As we will see, the curvature predicted by our temperature dependent
effective potential is smaller than the one corresponding to the
ground state PES, in the cases in which the quantum excited surfaces
approach, at the barrier top, the ground state one. Therefore, $T_0$
would be smaller than that corresponding to the ground state PES (see
Eq. (8) in~\cite{bothma-2009}), and hence the conclusion in
this reference ``that quantum tunneling does not play a significant
role in hydrogen transfer in enzymes'' is reinforced by our results.


\section{Method}
 We start our discussion with a system of
\emph{classical} particles, which we divide into two subgroups, one of
them, of mass $m$, characterized by the conjugate variables $(x,p)$,
and the other one, of mass $M$, characterized by the conjugate
variables $(X,P)$. If we had only one degree of freedom for each
subgroup (i.e. only one particle in one dimension), the Hamiltonian
would read:
\begin{equation}
\label{eq:classical-hamiltonian}
H_{\rm total}(x,p,X,P) = \frac{p^2}{2m}+\frac{P^2}{2M} + V_{\rm total}(x,X)\,.
\end{equation}
The generalization of the following to more particles or degrees of
freedom is straightforward. In the canonical ensemble, the average of any observable $O(x,p,X,P)$
is computed as:
\begin{equation}
\langle O(x,p,X,P)\rangle = \frac{1}{\mathcal{Z}} \int {\rm d}x{\rm
  d}p{\rm d}X{\rm d}P \; O(x,p,X,P) {\rm e}^{-\beta H_{\rm total}(x,p,X,P)}\,,
\end{equation}
where $\mathcal{Z} = \int {\rm d}x{\rm d}p{\rm d}X{\rm d}P {\rm
  e}^{-\beta H_{\rm total}(x,p,X,P)}$ is the partition function.  However, if we
think of an observable $A(X,P)$ that depends only on the variables of
one of the particles (let us call it a \emph{heavy} particle, assuming
that $M \gg m$), these two equations can be exactly rewritten as:
\begin{equation}
\langle A(X,P)\rangle = 
\frac{1}{\mathcal{Z}} \int {\rm d}X{\rm d}P A(X,P) {\rm e}^{-\beta H_{\rm eff}(X,P;\beta)}\,,
\end{equation}
with $\mathcal{Z} = \int {\rm d}X{\rm d}P {\rm e}^{-\beta H_{\rm eff}(X,P;\beta)}$.
Here, we have defined a temperature-dependent \emph{effective} Hamiltonian:
\begin{equation}
\label{Heff-classical}
H_{\rm eff}(X,P;\beta) := -\frac{1}{\beta}\log{
\int {\rm d}x{\rm d}p {\rm e}^{-\beta H_{\rm total}(x,p,X,P)}
}\,.
\end{equation}
Therefore, the equilibrium properties of the subsystem formed by the
heavy particle can be described in a closed manner, with an effective
Hamiltonian in which the coordinates of the light particle have been
integrated out.

If we now consider the two particles to be quantum,
the system will be characterized by the
canonical operators $(\hat{x},\hat{p},\hat{X},\hat{P})$, related by
the commutation relations, $[\hat{X},\hat{P}] = i\hbar\,,\;[\hat{x},\hat{p}]=i\hbar\,$,
and by a total Hamiltonian $\hat{H}_{\rm
  total}(\hat{x},\hat{p},\hat{X},\hat{P})$ whose expression is
analogous to Eq.~(\ref{eq:classical-hamiltonian}), except that we now
have operators instead of classical variables.

The key object, as far as equilibrium properties are concerned, is the equilibrium density matrix, defined as:
\begin{equation}
\hat{\rho}_{\rm eq} = \frac{1}{\mathcal{Z}} {\rm e}^{-\beta \hat{H}(\hat{x},\hat{p},\hat{X},\hat{P})}\,,
\;\;
\mathcal{Z} = {\rm Tr}[{\rm e}^{-\beta \hat{H}(\hat{x},\hat{p},\hat{X},\hat{P})}]\,,
\end{equation}
which allows to compute equilibrium averages as:
\begin{equation}
\langle \hat{O}(\hat{x},\hat{p},\hat{X},\hat{P})\rangle = {\rm Tr}[\hat{O}(\hat{x},\hat{p},\hat{X},\hat{P})\hat{\rho}_{\rm eq}]\,.
\end{equation}

Like in the classical case discussed before, for observables depending
only on the operators $\hat{X},\hat{P}$, the effective Hamiltonian can
be defined analogously to Eq.~(\ref{Heff-classical}) reading:
\begin{equation}
\label{Heff-quantum}
H_{\rm eff}(\hat{X},\hat{P};\beta) := -\frac{1}{\beta}\log{
{\rm Tr}_{\rm q} \left [ {\rm e}^{-\beta H_{\rm
      total}(\hat{x},\hat{p},\hat{X},\hat{P})} \right ]
}\,.
\end{equation}
Notice that the integrals in Eq.~(\ref{Heff-classical}) are now replaced
by the trace operation over the quantum Hilbert space 
denoted by ${\rm Tr}_{\rm q}$.


We are interested, however, in an intermediate case: the heavy
particle is classical, whereas the light particle is quantum
mechanical. For this purpose, it is most suitable to follow Kapral,
Nielsen and Ciccotti~\cite{kapral-1999, nielsen-2001} and start from
the partial Wigner representation~\cite{wigner-1932} of the full
quantum mechanical density matrix, $\hat{\rho}(t)$:
\begin{equation}
\hat{\rho}_{\rm W}(X,P,t) := (2\pi\hbar)^{-1} \int\!\!{\rm d}z\; e^{i Pz/\hbar}\langle X-z/2\vert \hat{\rho}(t) \vert X+z/2\rangle\,,
\end{equation}
which is an operator \emph{only} in the light particle Hilbert space,
depending on two real numbers $(X,P)$. The classical limit can then be
taken in a very straightforward manner by considering the evolution
equation for $\hat{\rho}_{\rm W}$, and retaining only linear terms
in $\sqrt{m/M} = \mu$. The result is~\cite{kapral-1999}:
\begin{equation}
\frac{\partial{\hat{\rho}}_{\rm W}}{\partial t}  =
-\frac{i}{\hbar}\left[ \hat{H}_{W, \rm total},\hat{\rho}_{\rm W}\right]
\label{eq:qcliouvillian}
+ \frac{1}{2}\left( 
  \lbrace \hat{H}_{W, \rm total},\hat{\rho}_{\rm W} \rbrace
- \lbrace \hat{\rho}_{\rm W}, \hat{H}_{W ,\rm total}  \rbrace
\right)\,.
\end{equation}
In this equation, the $\lbrace\cdot,\cdot\rbrace$ are the usual Poisson brackets.
The Hamiltonian $\hat{H}_{W, \rm total}(X,P)$ is  the partial
Wigner transform of the total Hamiltonian for the full quantum system
replacing the $(\hat{X},\hat{P})$ operators by real numbers:
\begin{equation}
\hat{H}_{W, \rm total}(\hat{x},\hat{p},X,P) = \frac{\hat{p}^2}{2m}+\frac{P^2}{2M} + V_{\rm total}(\hat{x},X)\,.
\end{equation}

The equilibrium density matrix in the partial Wigner representation at
the classical limit for the heavy particle, denoted by
$\hat{\rho}_W^{eq}$ should be stationary with respect to
Eq.~(\ref{eq:qcliouvillian}). If we use this property and expand it:
\begin{equation}
\hat{\rho}_W^{\rm eq} = \sum_{n=0}^{\infty} \mu^n \hat{\rho}_{W}^{\rm eq\;(n)}\,,
\end{equation}
it can then be proved~\cite{nielsen-2001} that the zero-th order term is given by:
\begin{eqnarray}
\label{eq:qcrho}
\hat{\rho}_{\rm W}^{\rm eq\;(0)} =  \frac{1}{\mathcal{Z}} {\rm
  e}^{-\beta \hat{H}_{W, \rm total}(\hat{x},\hat{p},X,P)}\,,
\end{eqnarray}
with 
$
\mathcal{Z} =  {\rm Tr}_{\rm q}\left[ \int {\rm d}X{\rm d}P {\rm
    e}^{-\beta\hat{H}_{W, \rm total}(\hat{x},\hat{p},X,P)}\right]\,.
$
Note that this object corresponds, at fixed classical variables
$(X,P)$, with the equilibrium density matrix \emph{for the electronic states}. 
However, it is only an \emph{approximation} to the true quantum-classical equilibrium
density matrix, since it is not a stationary solution to the
quantum-classical Liouvillian given in Eq.~(\ref{eq:qcliouvillian}).
The error made, i.e. the rate of change of the distribution as it evolves in time, is given by:
\begin{eqnarray}
\nonumber
\frac{\partial \hat{\rho}_W^{\rm eq(0)}}{\partial t} & = & 
\frac{P}{M}\frac{\beta}{\mathcal{Z}}\left[
\frac{1}{2}\left(
\frac{\partial \hat{H}_{W,{\rm total}}}{\partial X}e^{-\beta \hat{H}_{W,{\rm total}}} + e^{-\beta \hat{H}_{W,{\rm total}}}\frac{\partial \hat{H}_{W,{\rm total}}}{\partial X}
\right)
\right.
\\
& &
\left.
- \int_0^1 {\rm d}s e^{-\beta(1-s)\hat{H}_{W,{\rm total}}}\frac{\partial \hat{H}_{W,{\rm total}}}{\partial X}e^{-\beta \hat{H}_{W,{\rm total}}}
\right]
\end{eqnarray}
It can be seen that it grows with $\beta$ (quadratic dependence at
small $\beta$) and it is proportional to the velocity of the classical
particle $P/M$. Therefore, the error becomes smaller as the
temperature grows.

With these facts in mind, we will
consider Eq.~(\ref{eq:qcrho}) as a reasonable approximation and use it,
for example, to compute averages of observables:
\begin{eqnarray}
\nonumber
\langle \hat{O}(\hat{x},\hat{p},X,P)\rangle 
=  
{\rm Tr}_{\rm q}\int {\rm d}X{\rm d}P \hat{O}(\hat{x},\hat{p},X,P)
\hat{\rho}^{\rm eq \, (0)}_W \,.
\end{eqnarray}
Analogously to the preceeding sections [Cf. Eqs.~(\ref{Heff-classical}) and (\ref{Heff-quantum})],
if we are interested in observables that depend only on the heavy, classical particle,
we can define the temperature dependent effective Hamiltonian in the form:
\begin{equation}
\label{Heff-mix}
H_{\rm eff}(X,P;\beta) := -\frac{1}{\beta}\log{
{\rm Tr}_{\rm q}{\rm e}^{-\beta H_{\rm total}(\hat{x},\hat{p},X,P)}
}\,.
\end{equation}
It is then easy to verify that the partition function can be written
as:
\begin{equation}
\mathcal{Z} = \int {\rm d}X{\rm d}P {\rm e}^{-\beta H_{\rm eff}(X,P;\beta)}\,,
\end{equation}
and the classical observables can be computed as:
\begin{equation}
\label{averages}
\langle A(X,P)\rangle = \frac{1}{\mathcal{Z}}\int {\rm d}X{\rm d}P A(X,P) {\rm e}^{-\beta H_{\rm eff}(X,P;\beta)}\,.
\end{equation}
Hence, the quantum subsystem has been ``integrated out'', and does not
appear explicitly in the equations any more (of course, it has not
disappeared, being hidden in the definition of the effective
Hamiltonian).  In this way, the more complicated quantum-classical
calculations have been reduced to a simpler classical dynamics with an
appropriate effective Hamiltonian, which produces the same equilibrium
averages of classical observables [Eq.~(\ref{averages})] as the one we
would obtain using Eq.~(\ref{eq:qcrho}), and hence incorporates the
quantum back-reaction on the evolution of the classical variables



In the case of a molecular system, the total (partially Wigner
transformed) Hamiltonian reads:
\begin{equation}
\hat{H}_{\rm total}(R, P) = T_{\rm n}(P) + \hat{H}_{\rm e}(R)\,,
\end{equation}
where $R$ denotes collectively all nuclear coordinates, $P$ all
nuclear momenta, $T_n(P)$ is the total nuclear kinetic energy, and
$\hat{H}_{\rm e}(R)$ is the electronic Hamiltonian, that includes the
electronic kinetic term and all the interactions.
The effective Hamiltonian, defined in Eq.~(\ref{Heff-mix}) in general,
is in this case given by:
\begin{equation}
\label{eq:Heff}
H_{\rm eff}(R,P;\beta) = T_{\rm n}(P) - \frac{1}{\beta}\log{ {\rm Tr}_{\rm q} {\rm e}^{-\beta \hat{H}_{\rm e}(R)} }
:= T_{\rm n}(P) + V_{\rm eff}(R;\beta)\,,
\end{equation}
where the last equality is a definition for the \emph{effective}
potential.

Now, making use of the  adiabatic basis, defined as the set of all eigenvectors
of the \emph{electronic} Hamiltonian $\hat{H}_{\rm e}$:
$\hat{H}_{\rm e}(R)\vert\Psi_n(R)\rangle = E_n(R)\vert\Psi_n(R)\rangle
, $
we can rewrite $V_{\rm eff}(R;\beta)$ as:
\begin{equation}
\label{eq:veff}
V_{\rm eff}(R;\beta) = E_0(R) - \frac{1}{\beta}\log{ \left[1+\sum_{n>0}{\rm e}^{-\beta E_{n0}(R)}\right]}\,,
\end{equation}
where $E_{\rm n0}(R) = E_n(R)-E_0(R)$. This equation permits to see
explictly how the ground state energy $E_0$ differs from $V_{\rm
  eff}$, and in consequence how a MD based on $V_{\rm eff}$ is going
to differ from a gsBOMD. In particular, notice that $V_{\rm
  eff}(R;\beta) \le E_0(R)$.  Also, note that to the extent that
nuclei do not have quantum behavior near conical intersections or spin
crossings \cite{Yarkony}, nothing prevents us to use this equation
also in these cases.


The definition of the classical, effective Hamiltonian for the nuclear
coordinates in Eq.~(\ref{eq:Heff}) allows us now to use any of the
well-established techniques available for computing canonical
equilibrium averages in a classical
system~\footnote{\label{foot:mindT} Of course, since $H_{\rm eff}$ in
  Eq.~(\ref{eq:Heff}) depends on $T$, any Monte Carlo or dynamical
  method must be performed at the same $T$ that $H_{\rm eff}$ was
  computed in order to produce consistent results}, given in this case
by the convenient expression (\ref{averages}). For example, we could
use (classical) Monte Carlo methods, or, if we want to perform MD
simulations, we could propagate the stochastic Langevin dynamics
associated to the Hamiltonian (\ref{eq:Heff}):
\begin{equation}
\label{langevin-eq}
M_J\ddot{\vec{R}}_J(t) = -\vec{\nabla}_J V_{\rm eff}(R(t);\beta) - M_J\gamma\dot{\vec{R}}_J(t) + M_J\vec{\Xi}(t)\,,
\end{equation}
where $\vec{\Xi}$ is a vector of stochastic fluctuations, obeying 
$\langle \Xi_i(t)\rangle =  0 $ and 
$\langle \Xi_i(t_1)\Xi_j(t_2)\rangle =  2\gamma k_B T \delta_{ij}\delta (t_1-t_2)$
which relates the dissipation strength $\gamma$ and the temperature $T$
to the fluctuations (fluctuation-dissipation theorem).

Indeed, it is well-known that these dynamics are equivalent to the
Fokker-Planck equation for the probability density $W(R,P)$ in the
classical phase space \cite{VanKampen}:
\begin{equation}
\label{FP-gral}
 \frac{\partial W (R,P;t)}{\partial t}
=
\{H_{\rm eff}(R,P;\beta), W (R,P;t) \} 
+
\gamma 
\sum_J 
\partial_{\vec {P}_J} (\vec {P}_J  + M k_{\rm B} T \partial_{\vec {P}_J})  W(R,P;t)
\end{equation}
where $ \{ \cdot, \cdot \}$ is the classical Poisson bracket.

Any solution to Eq.~(\ref{FP-gral}) approaches at infinite time a distribution
$W_{\rm eq}(R,P)$ such that $\partial_t W_{\rm eq}(R,P) = 0$. This stationary
solution is unique and equal to the Gibbs distribution, $W_{\rm eq}(R,P) =
\mathcal{Z}^{-1} \; {\rm e}^{-\beta H_{\rm eff}(R,P;\beta)}$ \cite{VanKampen}.
Thus, the long time solutions of Eq.~(\ref{FP-gral}), and hence those of
Eq.~(\ref{langevin-eq}) reproduce the canonical averages in Eq.~(\ref{averages}). This
property, which is also satisfied by other dynamics like the one proposed by
Nos\'e \cite{nose1,nose2} if the $H_{\rm eff}$ in Eq.~(\ref{eq:Heff}) is used,
comes out in a very natural way from the present formalism while it is yet
unclear of other ab initio MD candidates for going beyond gsBOMD
\cite{mauri-1993, parandekar-2005, schmidt-2008, bastida-2007}.


\section{Applications}
 
The most obvious applications of our temperature dependent effective
potential and its associated dynamics are the processes of proton
transfer and thermal isomerization whenever low lying electronic
excitated states have to be taken into account. These processes are
ubiquitous in chemistry and biochemistry \cite{Bell, may-2004, berson}. In
particular, ab initio molecular dynamics of intramolecular proton
transfer around conical intersection and excited states is a topic of
current interest \cite{coe} and a great tradition \cite{douhal}.

\begin{figure}
\setlength{\unitlength}{\columnwidth}
\begin{picture}(1.0,0.7)
\put(0.00,0.70){\includegraphics[width=0.7\columnwidth,angle=270]{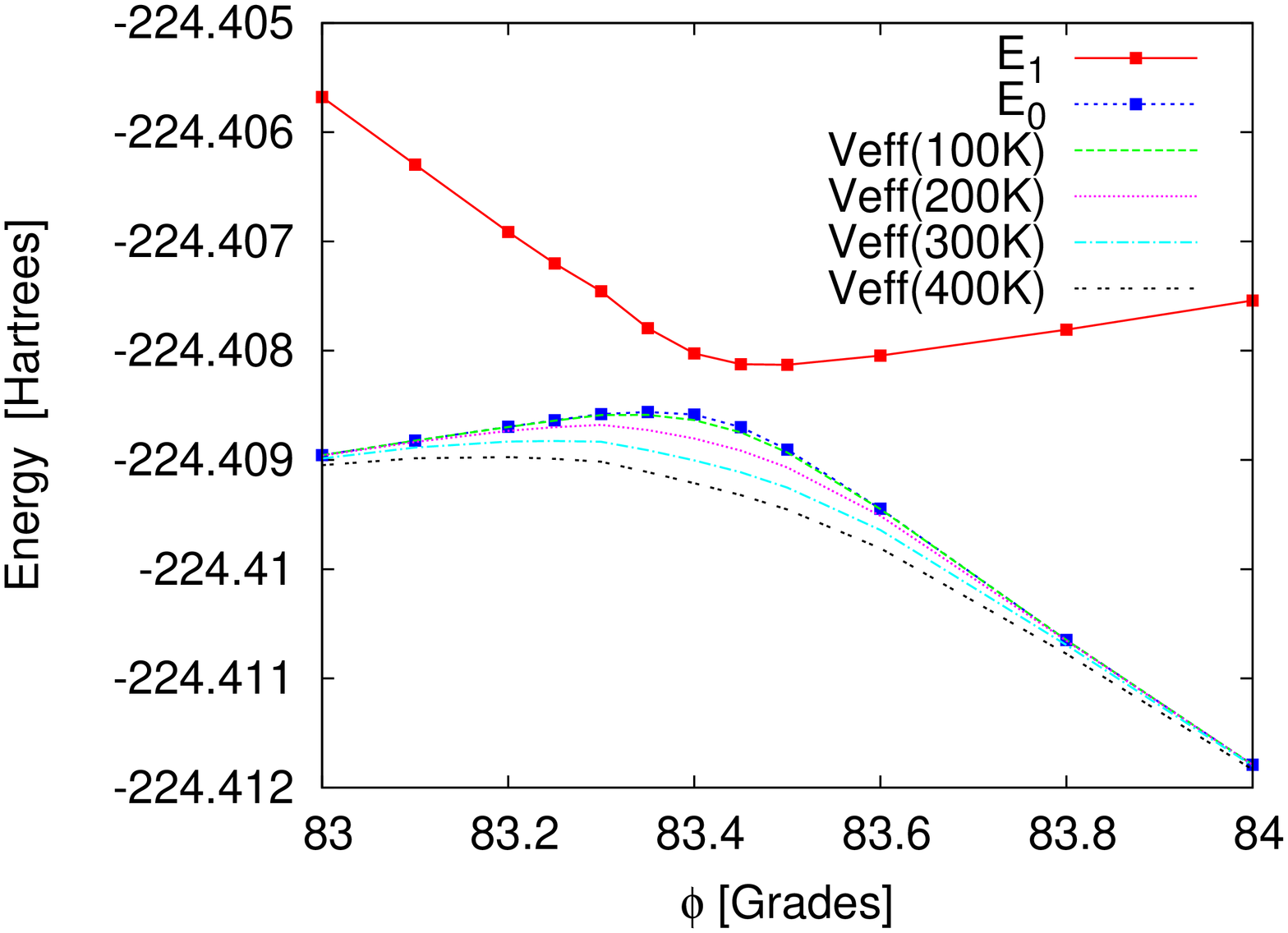}}
\put(0.33,0,15){\includegraphics[width=0.35\columnwidth]{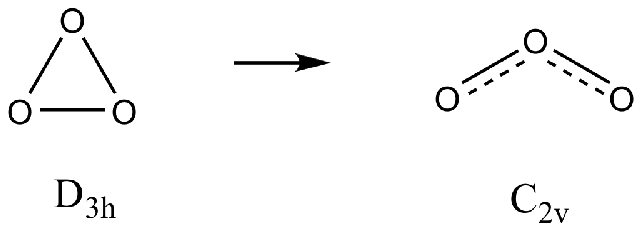}}
\end{picture}
\caption{
\label{fig1}
(color online)
PES corresponding to the $1^1A_1 (E_0)$ and $2^1A_1 (E_1)$ states, and
effective potential at the indicated temperatures, for the ozone
molecule. The reaction coordinate is the molecular angle.}
\end{figure}

 As an example of our approach, we show in the
following the difference between our temperature dependent effective
potential and the gs BO PES for the avoided crossing between the open
and closed forms of the ozone molecule. We focus on the ring closure
of this system, as depicted in Fig.~\ref{fig1} (inset).  Using the
CASSCF method (complete active space self consistent field)~\cite{roos-1987}, we have computed the PES corresponding
to the ground and first excited singlet states ($1^1A_1$ and
$2^1A_1$), and the effective potential along the relevant reaction
coordinate, the ring closure angle $\phi$ -- using only those states
as they are the only relevant ones in this case. At around $\phi_0 =
\phi=83.4^{\rm o}$, there is a barrier between two possible minima in
the gs PES of this molecule. The $2^1A_1$ electronic state approaches
at this point the ground state PES, and in consequence one might
expect that the effective potential, and its derivatives at the cusp,
will differ considerably from the gs PES values as the temperature
goes up. This can be seen in Fig.~\ref{fig1}.

The situation shown in this figure is very general. In fact, one can
prove from Eq.~\ref{eq:veff} that if the first and second derivative of $E_{10}$ at the
barrier top verifies:
\begin{equation}
\frac{\partial^2 E_{10}}{\partial \phi^2}(\phi_0) \left( 1+ e^{-\beta E_{10}(\phi_0)}\right)
> 
\beta \left( \frac{\partial E_{10}}{\partial \phi}(\phi_0) \right)^2\,,
\end{equation}
then:
\begin{equation}
\left| \frac{\partial^2 V_{\rm eff}}{\partial \phi^2(\phi_0;\beta)} \right|
<
\left| \frac{\partial^2 E_0}{\partial \phi^2}(\phi_0) \right|\,,
\end{equation}
i.e. the curvature by our $V_{\rm eff}$ is smaller than the gs PES curvature. Note
that, in avoided crossings $\frac{\partial E_{10}}{\partial \phi}(\phi_0)$ is
approximately zero  and therefore the above condition will be verified in the most
interesting cases.

Here, we have computed the effective potential energy
surface for a given reaction coordinate. Once this surface is our
disposal, it is a trivial task to perform nuclear dynamics on top of
this surface. As we mentioned in the previous section, this nuclear
dynamics can be performed in two ways. First of all, one could perform
dynamics for ensembles with the Fokker-Planck equation. In this case,
one would just use the effective potential that has been pre-computed
in the equation, and use any of the well-known partial differential
equation solvers to propagate Fokker-Planck's equation. Alternatively,
one can propagate the nuclear dynamics individually,
Eq. (\ref{langevin-eq}),  and then one would use  any of
the standard propagators that are also well described in the
literature.

However, in many situations it will not be advantageous to pre-compute
the effective potential energy surface (due to a larger
dimensionality, etc), and instead the propagation would be peformed 
"on the fly", i.e. simultaneously to the computation of only those
points in the surface that are necessary.

Note that the computational cost for the evaluation of the temperature
dependent effective potential is necessarily larger than the cost for
the computation of the gsBOPES. A number of excited state surfaces
must be computed, which can be done, for example, with the CASSCF 
technique that we have employed here, or with time-dependent
density-functional theory. If we wish to perform 'on the fly' MD
simulations, we will also need to compute the
gradients of those surfaces, which can be done with linear response
theory -- similarly to how it is done for the ground state
surface.

\section{Conclusions}
 In this work  we have introduced a temperature
dependent effective potential and the associated constant-$T$ dynamics
which produce in a natural way the Boltzmann equilibrium population of
the quantum subsystem and its corresponding back reaction; something
pursued in recent years by many researchers~\cite{mauri-1993,
  parandekar-2005, schmidt-2008, bastida-2007}. We justify our only
assumption, the equilibrium distribution prescribed by
Eq.~(\ref{eq:qcrho}), using the formulation of Nielsen, Kapral and
Ciccotti~\cite{kapral-1999,nielsen-2001}. Our approach is particularly
useful in the case of conical intersection or spin-crossing
\cite{Yarkony}, and does not assume that the electrons or quantum
variables immediately follow the nuclear motion, in contrast to any
adiabatic approach. The fact that, when using our effective potential,
the height of a barrier in the PES and its curvature near the top
decrease at avoided crossings makes our work relevant in the context
of the transition-state theory \cite{HTB}. In particular, this is
important if one wants to adequately discriminate possible
quantum-like behavior of nuclei from simple classical effects due to
the direct influence of temperature on the potential between two
metastable states, for example in biological
systems~\cite{bothma-2009,iyengar-2008}.

\section*{Acknowledgements}
  We gratefully thank Giovanni Ciccotti, Roberto Car,  Fernando
  Falceto and Karsten Reuter for their
  valuable comments.  We acknowledge support by the Spanish MICINN
  (FIS2007-65702-C02-01, FIS2009-13364-C02-01, FIS2008-01240 and ACI-Promociona
  ACI2009-1036), by ``Grupos de Excelencia del Gobierno de Arag\'on''
  (E24/3) by ``Grupos Consolidados UPV/EHU del Gobierno Vasco''
  (IT-319-07), by the CSIC (200980I064) by the European Union through e-I3, 
  and by the ETSF project (Contract Number 211956).


\end{document}